%% file: main.tex
\documentclass[sigconf]{acmart}

\usepackage{balance}

\usepackage{xspace}
\usepackage{subcaption}
\usepackage{multirow}
\usepackage{enumitem}
\usepackage{listings}
\usepackage{makecell}
\usepackage{colortbl}

\makeatletter
\AtBeginDocument{%
  \begingroup
  \small
  \let\tmp@n@s\f@size
  \let\tmp@n@b\f@baselineskip
  \normalsize
  \let\tmp@s@s\f@size
  \let\tmp@s@b\f@baselineskip
  \xdef\semismall@size{\fpeval{(\tmp@n@s+\tmp@s@s)/2}}%
  \xdef\semismall@baselineskip{\fpeval{(\tmp@n@b+\tmp@s@b)/2}}%
  \endgroup
}
\newcommand{\semismall}{\fontsize{\semismall@size}{\semismall@baselineskip}\selectfont}

\newcommand{\toolName}{Caption\xspace}

\AtBeginDocument{%
  }

\setcopyright{acmlicensed}
\copyrightyear{2025}
\acmYear{2025}
\setcopyright{rightsretained}
\acmConference[UIST Adjunct '25]{The 38th Annual ACM Symposium on User Interface Software and Technology}{September 28-October 1, 2025}{Busan, Republic of Korea}
\acmBooktitle{The 38th Annual ACM Symposium on User Interface Software and Technology (UIST Adjunct '25), September 28-October 1, 2025, Busan, Republic of Korea}\acmDOI{10.1145/3746058.3758413}
\acmISBN{979-8-4007-2036-9/2025/09}

\begin{document}

\title{\toolName: Generating Informative Content Labels for Image Buttons Using Next-Screen Context}

\author{Mingyuan Zhong}
\orcid{0000-0003-3184-759X}
\affiliation{%
  \institution{University of Washington}
  \city{Seattle}
  \state{WA}
  \country{USA}
}
\email{myzhong@cs.washington.edu}

\author{Ajit Mallavarapu}
\orcid{0009-0002-5358-4073}
\affiliation{%
  \institution{University of Washington}
  \city{Seattle}
  \state{WA}
  \country{USA}
}
\email{ajitm@uw.edu}

\author{Qing Nie}
\orcid{0009-0002-4757-5090}
\affiliation{%
  \institution{University of Washington}
  \city{Seattle}
  \state{WA}
  \country{USA}
}
\email{qngnie@gmail.com}


\begin{abstract}

We present \textit{\toolName}, an LLM-powered content label generation tool for visual interactive elements on mobile devices. 
Content labels are essential for screen readers to provide announcements for image-based elements, but are often missing or uninformative due to developer neglect.
Automated captioning systems attempt to address this, but are limited to on-screen context, often resulting in inaccurate or unspecific labels. To generate more accurate and descriptive labels, \toolName collects next-screen context on interactive elements by navigating to the destination screen that appears after an interaction and incorporating information from both the origin and destination screens.
Preliminary results show \toolName generates more accurate labels than both human annotators and an LLM baseline.
We expect \toolName to empower developers by providing actionable accessibility suggestions and directly support on-demand repairs by screen reader users.

\end{abstract}




\maketitle

\section{Introduction}
    \input{sections/10-introduction}

\section{\toolName: System Design}\label{sec:system}
    \input{sections/20-system}

\section{Preliminary Evaluations}
    \input{sections/30-eval}

\section{Discussion and Future Work}
    \input{sections/40-discussion}


\bibliographystyle{ACM-Reference-Format}
\bibliography{ref}










\end{document}

%% file: sections/10-introduction.tex
Content labels are text alternatives for visual UI elements, essential for people who use screen readers to access their devices.
Although required by industry~\cite{w3c_wcag2.2, appt_guidelines, wild2023mobile} and platform guidelines~\cite{material_design_guidelines, apple_guidelines, microsoft_guidelines}, many mobile apps have been found to be inaccessible due to missing or inaccurate labels~\cite{fok_large-scale_chi22, mateus2020eval, ross_examining_assets18}.
To address this issue, crowd annotators were recruited to create labels for image-based elements to train machine learning models~\cite{li2020widget, zhang_screen_chi21}.
Some of the resulting models have been deployed in mobile operating systems as part of their built-in screen readers~\cite{google_whats_nodate, apple_use_2023}.
However, these annotation efforts are costly, have difficulty in covering less common elements, and are prone to receiving low-quality or inconsistent labels.

\begin{figure}
  \includegraphics[width=0.47\textwidth]{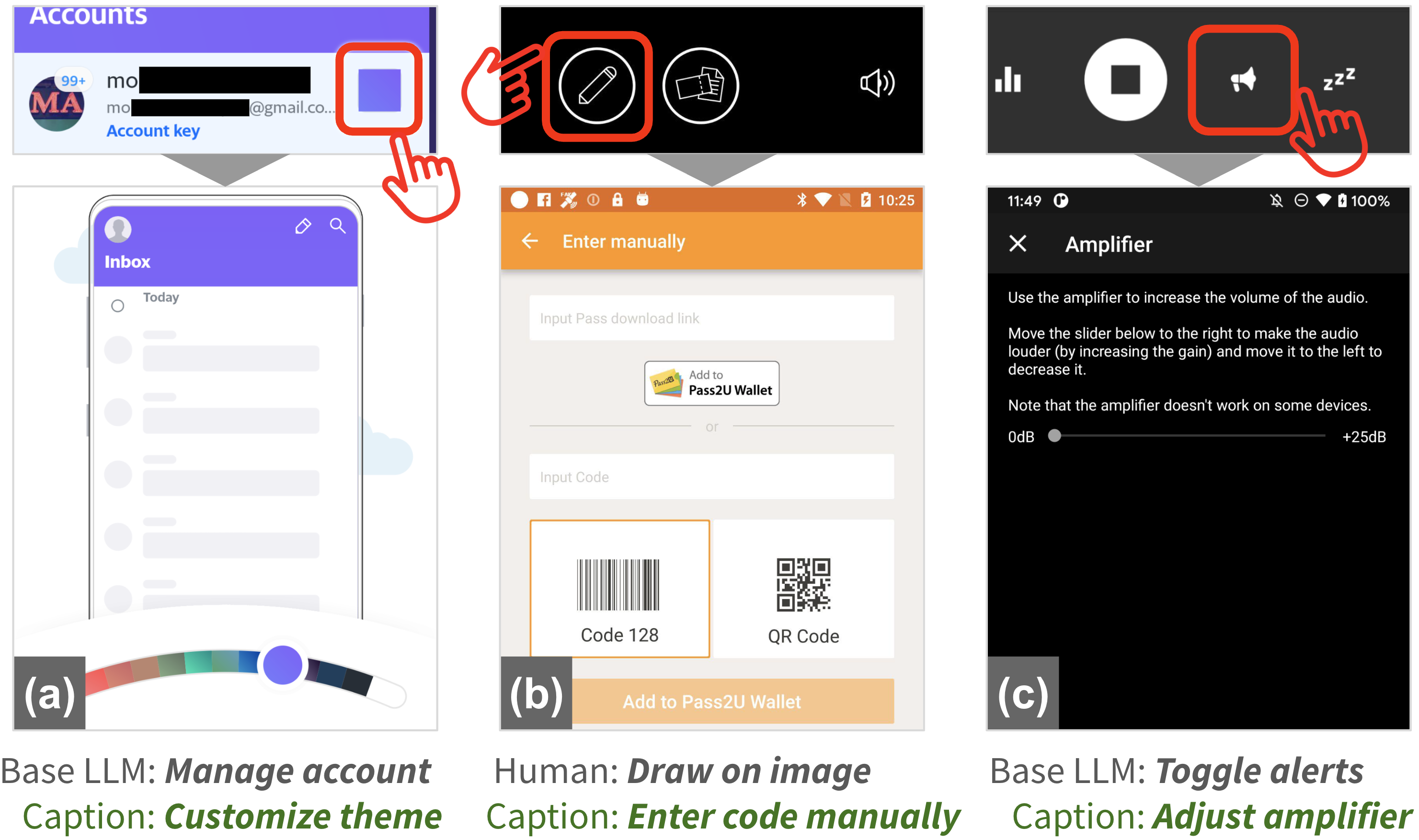}
  \caption{Buttons (circled) can be difficult to annotate without destination screen context (bottom). Unlike an LLM baseline and human annotators, Caption generated accurate labels.}
  \label{fig:examples}
  \Description{Examples of mobile application buttons with labels for comparison.
(a) A square, purple button next to a user profile, which leads to a color picker screen. Base LLM’s label is “Manage account”. Caption’s label is “Customize theme”.
(b) A pencil button in a toolbar, which leads to a screen that allows manual entry of a code or download link. Human annotator's label is “Draw on image”. Caption’s label is “Enter code manually”.
(c) A loudspeaker button in a playback toolbar, which leads to a screen with an amplifier slider. Base LLM’s label is “Toggle alerts”. Caption’s label is “Adjust amplifier”.}
  \vspace{-9px}
\end{figure}

Recent advances in LLM-driven UI understanding have enabled tools that automatically detect nuanced accessibility errors~\cite{taeb24axnav, zhong2025screenaudit} and generate repairs~\cite{liu2024Unblind, othman2023fostering, huang2024access}.
Despite current LLMs' capabilities, existing automated captioning techniques rely on contexts \textit{within} the same screen as the element being captioned~\cite{zhang_screen_chi21, baechler2024screenai, lu2024omniparser, wang2021screen2words}.
This can lead to inaccuracies in labels when necessary context does not exist.
Figure \ref{fig:examples} shows three such examples, where the buttons' purposes only became clear after viewing the destination screens.

\textit{\toolName} aims to improve the description quality for interactive elements by including additional context from the next screen.
Our preliminary results show improvements in label quality for image-based buttons, especially when buttons use generic pictograms (Figure~\ref{fig:examples}\,a) or when the action triggered by a button does not strictly align with what the button's appearance implies  (Figure~\ref{fig:examples}\,b,c).

%% file: sections/20-system.tex
We designed \toolName to generate high-quality content labels by considering 
(1) the interactive button itself,
(2) the surrounding context on screen, and
(3) the destination screen resulting from clicking the button.
\toolName explores the destination screen by programmatically clicking on the button of interest, then captures the destination screen after a predetermined timeout.
This exploration process is similar to existing automated crawling techniques in security~\cite{hao_puma_mobisys14}, interface design~\cite{deka_erica_uist16, deka_rico_uist17}, and accessibility~\cite{fok_large-scale_chi22}.

In this paper, we focus on the design of an effective captioning pipeline after collecting the necessary screenshot, metadata, and destination screen information.
Current interface captioning tools either directly use LLMs for generation~\cite{song2025altauthor, pedemonte2025improving, huang2024access} or fine-tune smaller models on LLM labels~\cite{lu2024omniparser, baechler2024screenai, chai2024amex}.
In all these efforts, only the buttons and current screen contexts are available to the models.

\toolName additionally leverages the destination screen to improve the quality of generated content labels.
We explore three prompting designs that incorporate the destination screen:
(1)~include only the destination screenshot,
(2)~include only an LLM-generated destination screen description, and
(3)~include an LLM-generated description and screenshot of the destination screen.
We developed the prompts by following strategies from prior work~\cite{lu2024omniparser, chai2024amex} and referencing accessibility guidelines~\cite{w3c_wcag2.2, material_design_guidelines}.
When generating the description for the destination screen, we instructed the LLM to ignore navigational features (e.g., back buttons, tab bars), and use generic description for dynamic contents (e.g., news articles).

%% file: sections/30-eval.tex
We conducted preliminary evaluation of \toolName on three existing Android crawl datasets: RICO (2017)~\cite{deka_rico_uist17}, Fok et al.'s (2022)~\cite{fok_large-scale_chi22}, and AMEX (2024)~\cite{chai2024amex}.
Each of the datasets contains the following types of data:
screenshots,
metadata (e.g., button classes, locations, developer assigned content labels),
and interaction traces.
From each dataset, we randomly sampled 160 image-based buttons with associated destination screen information based on available metadata and interaction traces, with a total of 480 buttons.

We carried out two evaluations: (1) a prompt analysis that aims to understand whether the prompting strategies are meaningfully different (using 78 examples) and 
(2) a system evaluation that compares \toolName generated content labels against baselines (using the other 402 examples).
We used Google's Gemini 2.5 Flash.

\begin{figure}
  \includegraphics[width=0.35\textwidth]{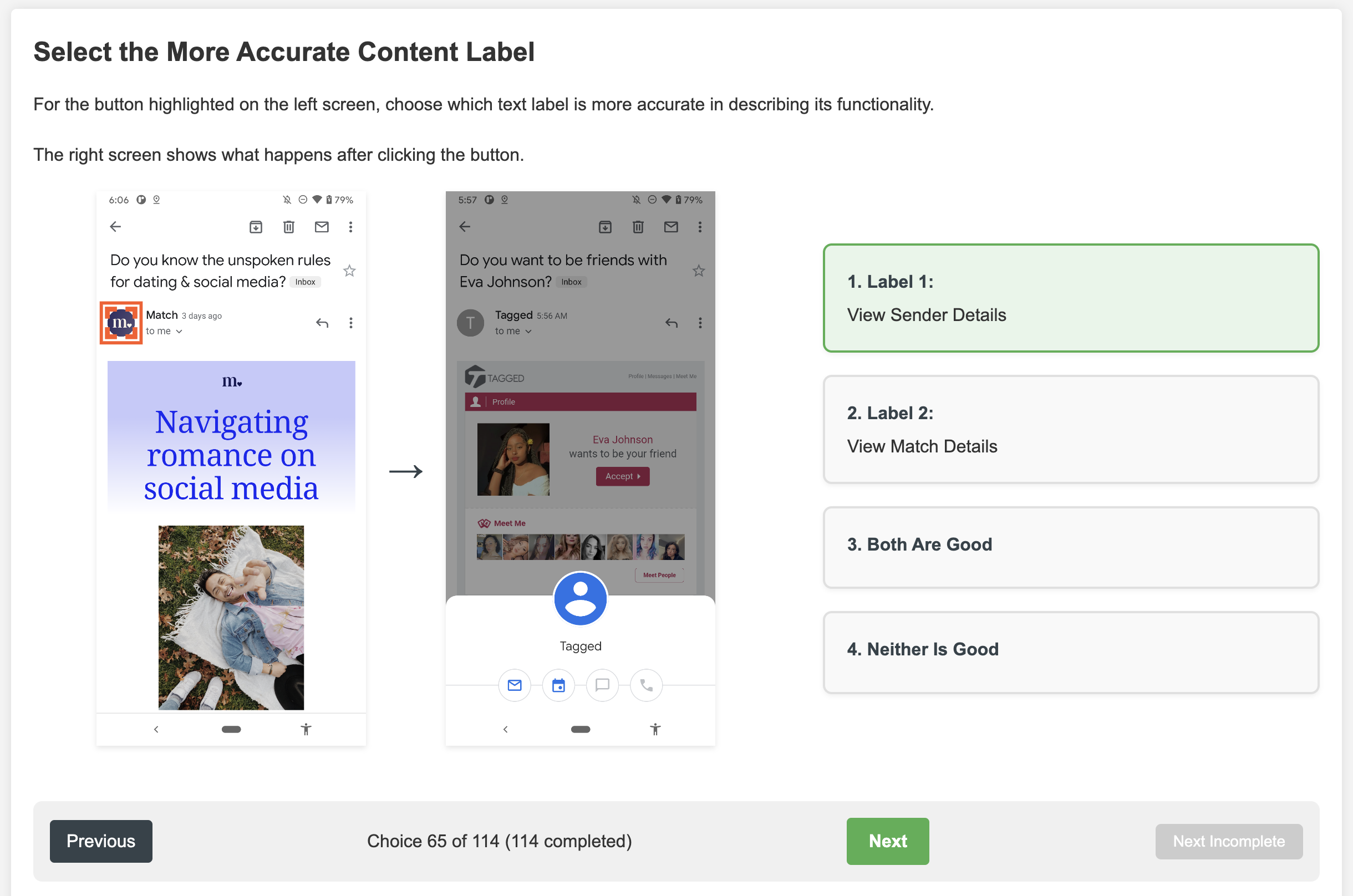}
  \caption{Label selection interface used in the evaluations.}
  \label{fig:annotation}
  \Description{Screenshot of the label selection interface. The interface presents a button’s screenshot and multiple label options. Raters can select their preferred label, choose ‘both’ if equally good, or ‘neither’ if both are inadequate.}
  \vspace{-10pt}
\end{figure}

Seven raters (3 women, 4 men, ages 20--33) participated in the evaluation, using the interface as shown in Figure~\ref{fig:annotation}. 
In particular, raters chose ``neither'' when encountering unclear actions or highlighting errors (e.g., from data discrepancies). 
All presentation was randomized.
Raters each evaluated between 180 and 240 buttons, with each session lasting about one hour.
To assess inter-rater reliability, we computed Cohen's $\kappa$ on the ratings.

\subsection{Results}
\subsubsection{Prompt Analysis}
We compared the three prompting strategies described in Section~\ref{sec:system}.
In total, 468 pairwise comparisons were collected, with each pair of labels receiving two raters' choices.
We manually examined all examples that received a ``neither'' rating as these may indicate low data quality.
We found 15 examples (19.2\%) to be problematic (i.e., displayed an implausible transition or highlighted a non-interactive part of screen) and excluded them.
Agreement between raters was moderate, with Cohen's $\kappa=0.402$.

An analysis of variance based on logistic regression~\cite{berkson1944application} found no detectable effect of \textit{Prompt} on \textit{Preference}, $\chi^2(2,N\mathord{=}756) = 0.96$, $p = .62$.
Therefore, we consider all three prompting strategies with next screen contexts to produce labels of similar quality.

\subsubsection{System Evaluation}
We compared the content labels that \toolName generated against two baseline techniques:
(1) Human annotations from Widget Captioning~\cite{li2020widget} that covered RICO (134 examples) and
(2) LLM-generated baseline labels based on AMEX's prompts~\cite{chai2024amex} and accessibility guidelines~\cite{w3c_wcag2.2, material_design_guidelines}, applied to all three datasets (402 examples).
In total, 1,072 [$=(134+402)\times2$] pairwise comparisons were evaluated, with each pair of labels receiving two raters' choices.
We manually examined all examples that received a ``neither'' rating and excluded 43 examples (10.7\%) from analysis.
Agreement between raters was moderate, with Cohen's $\kappa=0.420$.

\begin{figure}
  \includegraphics[width=0.4\textwidth]{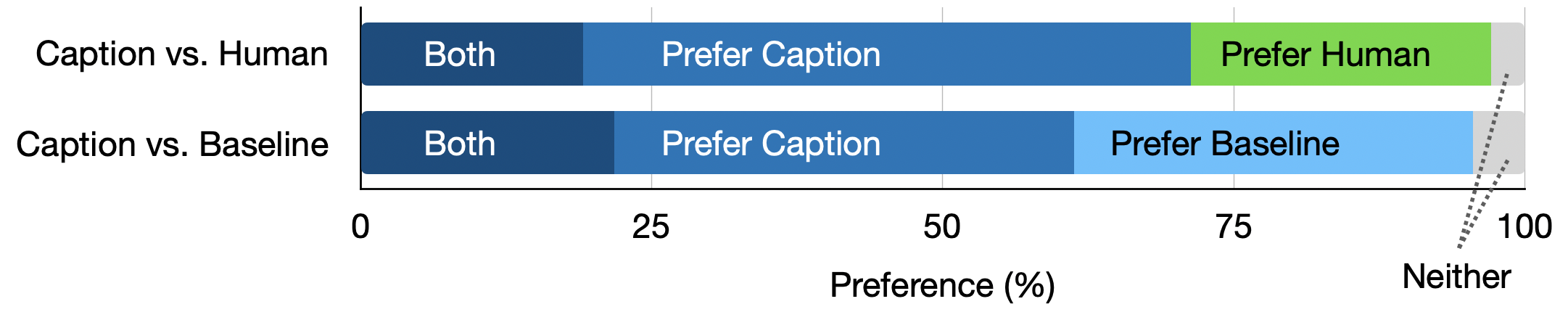}
  \caption{Rater preferences in our system evaluation.}
  \label{fig:preference}
  \Description{Bar chart showing rater preferences in system evaluation for Caption compared with Human annotations and with LLM Baseline. In the Caption vs. Human comparison, 19.12\% preferred both equally, 52.21\% preferred Caption, 25.74\% preferred the LLM Baseline, and 2.94\% preferred neither. In the Caption vs. Baseline comparison, 21.77\% preferred both, 39.52\% preferred Caption, 34.27\% preferred Human annotations, and 4.44\% preferred neither. The results indicate Caption was preferred over both baselines in the majority of cases.}
  \vspace{-10pt}
\end{figure}

An analysis of variance based on logistic regression~\cite{berkson1944application} indicated a statistically significant effect of \textit{Technique} on \textit{Preference}, $\chi^2(2,N\mathord{=}1852) = 37.2$, $p < .001$.
\textit{Post hoc} pairwise comparisons, corrected with Holm’s sequential Bonferroni procedure~\cite{holm_bonferroni}, indicated that \toolName vs. LLM Baseline ($Z=2.23, p=.026$) and \toolName vs. Human ($Z=5.11, p<.001$) were significantly different.
Figure~\ref{fig:preference} shows rater preference distribution.
When compared to human annotations, raters preferred \toolName 52.2\% of the time and preferred human annotations 25.7\% of the time.
When compared to the LLM baseline, raters preferred \toolName 39.5\% of the time and preferred the LLM baseline 34.3\% of the time.

%% file: sections/40-discussion.tex
Our preliminary results show that \toolName enables the generation of more accurate  content labels for image-based buttons by incorporating destination screen contexts, outperforming both an LLM baseline using on-screen context and human annotations.
We anticipate \toolName to be useful in suggesting high-quality content labels for developers, as well as supporting screen reader users when navigating currently inaccessible apps.

While our analyses covered multiple existing datasets, we operated on a small subset of data to explore and validate our strategy.
Future work should further evaluate the technique using larger datasets and involve people who regularly use screen readers.
We also see opportunities for adopting \toolName in various application scenarios.
\toolName-generated informative content labels can be directly adopted using an interaction proxy approach~\cite{zhang_interaction_chi17} to support runtime repairs of inaccessible image-based elements.
\toolName can also be used to generate more descriptive and accurate labels for training on-device UI understanding models, similar to prior work~\cite{zhang_screen_chi21, lu2024omniparser}.
These models are foundational for the reliable execution of complex tasks, supporting agentic task execution~\cite{vu2024gptvoicetasker, wang2024mobileagentv2mobiledeviceoperation} and usability evaluation~\cite{taeb24axnav, xiang2024simuser}.
